\documentclass[sigconf,natbib=true]{acmart}

\usepackage{booktabs}
\usepackage{enumitem}
\usepackage{physics}
\usepackage{tabularx}
\usepackage{amsmath}
\usepackage{subcaption}
\usepackage{siunitx}
\usepackage{gensymb}
\usepackage{hyperref}
\usepackage{balance}
\hypersetup{
    colorlinks=true,
    linkcolor=blue,
    filecolor=black,
    urlcolor=blue,
    citecolor=blue
}

\copyrightyear{2025}
\acmYear{2025}
\setcopyright{cc}
\setcctype{by}
\acmConference[SIGIR '25]{Proceedings of the 48th International ACM SIGIR Conference on Research and Development in Information Retrieval}{July 13--18, 2025}{Padua, Italy}
\acmBooktitle{Proceedings of the 48th International ACM SIGIR Conference on Research and Development in Information Retrieval (SIGIR '25), July 13--18, 2025, Padua, Italy}\acmDOI{10.1145/3726302.3730065}
\acmISBN{979-8-4007-1592-1/2025/07}

\begin{document}
\title{QDER: Query-Specific Document and Entity Representations for Multi-Vector Document Re-Ranking}

\author{Shubham Chatterjee}
\orcid{1234-5678-9012}
\affiliation{%
    \institution{Missouri University of Science and Technology} 
    \department{Department of Computer Science} 
    \city{Rolla} 
    \state{Missouri} 
    \country{United States}
}
\email{shubham.chatterjee@mst.edu}

\author{Jeff Dalton}
\orcid{0000-0003-2422-8651}
\affiliation{%
    \institution{The University of Edinburgh} 
    \department{School of Informatics} 
    \city{Edinburgh} 
    \state{Scotland} 
    \country{United Kingdom}
}
\email{jeff.dalton@ed.ac.uk}



\begin{abstract}
 
Neural IR has advanced through two distinct paths: entity-oriented approaches leveraging knowledge graphs and multi-vector models capturing fine-grained semantics. We introduce \texttt{QDER}, a neural re-ranking model that unifies these approaches by integrating knowledge graph semantics into a multi-vector model.

\texttt{QDER}'s key innovation lies in its modeling of query-document relationships: rather than computing similarity scores on aggregated embeddings, we maintain individual token and entity representations throughout the ranking process, performing aggregation only at the final scoring stage—an approach we call ``late aggregation.'' We first transform these fine-grained representations through learned attention patterns, then apply carefully chosen mathematical operations for precise matches. Experiments across five standard benchmarks show that \texttt{QDER} achieves significant performance gains, with improvements of 36\% in nDCG@20 over the strongest baseline on TREC Robust 2004  and similar improvements on other datasets. \texttt{QDER} particularly excels on difficult queries, achieving an nDCG@20 of 0.70 where traditional approaches fail completely (nDCG@20 = 0.0), setting a foundation for future work in entity-aware retrieval.
\end{abstract}

\maketitle

\section{Introduction}
\label{sec:Introduction}

\textbf{Background.} Information retrieval (IR) systems face a fundamental challenge: effectively modeling the complex semantic relationships between queries and documents. Traditional retrieval methods relying on lexical matching often suffer from vocabulary mismatch and fail to capture deeper semantic connections. These challenges are particularly acute for complex queries that require understanding specific entity relationships and broader contextual relevance. For example, answering a query like ``\textit{What was Steve Jobs' role in the development of the iPhone?}'' requires not only recognizing mentions of ``Steve Jobs'' and ``iPhone'' but also accurately interpreting their relationship, rather than treating them as independent concepts.

One promising direction has been the development of entity-oriented neural IR models, which enrich text representations with knowledge from Knowledge Graphs (KGs). The Entity-Duet Ranking Model \cite{liu-etal-2018-entity} follows this approach by deriving entity vectors from KG descriptions and types, and matching them to queries using a word-entity duet framework \cite{xiong2017word}. Tran and Yates \cite{tran2022dense} organized document entities into clusters, providing multiple perspectives that improved document understanding and retrieval performance.

In parallel, neural IR has shifted toward multi-vector document representations that maintain token-level embeddings for fine-grained matching between query and document components. Models such as ColBERT \cite{khattab2020colbert} and ME-BERT \cite{luan-etal-2021-sparse} exemplify this approach. We hypothesize that multi-vector representations are particularly beneficial for entity-oriented search, where preserving distinct entity contexts can significantly improve retrieval accuracy. For example, a query like ``\textit{What was Steve Jobs' role in the development of the iPhone?}'' may be better handled by multi-vector models that separate Jobs' various roles across companies, whereas single-vector models risk conflating them.

These advances raise a fundamental research question: \textit{Can the integration of entity representations into multi-vector neural frameworks further improve retrieval performance?} In this work, we explore this direction, proposing a model that unifies entity-aware and fine-grained matching capabilities. We focus on the re-ranking phase of retrieval systems. Given a query $Q$ and a set of candidate documents $\mathcal{D}$ retrieved by an initial retriever (e.g., BM25 \cite{robertson2009probabilistic}), the goal is to \textit{re-rank} $\mathcal{D}$ according to relevance to $Q$, using computationally intensive but more precise matching.

\textbf{Research Gap.} Current neural IR models face two key limitations. First, traditional matching methods based on dot product or cosine similarity \cite{karpukhin-etal-2020-dense,reimers2019sentencebertsentenceembeddingsusing,bingyu-arefyev-2022-document,khattab2020colbert,xiong2021approximate} often fail to capture the complex semantic relationships required for accurate relevance assessment. We hypothesize that effective retrieval demands modeling how different components of query and document embeddings interact, as each encodes distinct semantics. Second, most systems rely on static, query-agnostic document representations \cite{kulkarni2023ladr,liu-etal-2018-entity,khattab2020colbert,macavaney2019ceder}. However, prior work \cite{chatterjee2022berter,chatterjee2024dreq} shows that such static embeddings significantly underperform on nuanced, multifaceted queries.

To address these challenges, we propose \textit{\textbf{Q}uery-specific \textbf{D}ocument and \textbf{E}ntity \textbf{R}epresentations} (\texttt{QDER}), a novel entity-oriented multi-vector neural re-ranking model.\footnote{\textbf{Github: \url{https://github.com/shubham526/SIGIR2025-QDER}.}} At its core, \texttt{QDER} treats documents as collections of tokens and entities, deriving distinct representations for each. While prior work has analyzed documents through both textual and entity lenses \cite{liu-etal-2018-entity,xiong2016bag,xiong2017word}, our key innovation lies in making these representations \emph{query-specific}. 

\texttt{QDER} employs two complementary mechanisms to generate query-specific representations. First, it uses attention to dynamically reweight document tokens and entities based on their relevance to the query, transforming static document representations into query-focused ones. Second, building on research showing that elements within relevant documents are themselves often relevant \cite{nguyen2016marco,dietz2017trec}, we incorporate initial retrieval scores to further refine these representations. Together, these mechanisms highlight the document elements most pertinent to the query while mitigating noise from imperfect entity linking. Through this integration of fine-grained multi-vector modeling, entity knowledge, and query-specific dynamic refinement, \texttt{QDER} establishes a new state-of-the-art for entity-aware neural re-ranking.

\emph{\textbf{Our model introduces several key innovations}}:
\begin{itemize}[leftmargin=*]
   \item \textbf{Late Aggregation:} Building on the late interaction idea from ColBERT \cite{khattab2020colbert}, we propose \emph{late aggregation}. We preserve and jointly model token and entity representations throughout the ranking process, and aggregate them only at final scoring.

    \item \textbf{Attention-guided Query-specific Representations:} We extend the late interaction paradigm by introducing \emph{entity-aware attention that dynamically reshapes token and entity representations based on query context.} These refined representations undergo addition and multiplication with query embeddings to model complementary and exact semantic matches. Unlike previous work that relied solely on token matching or static entity characteristics \cite{xiong2017end,liu-etal-2018-entity,tran2022dense}, our unified framework preserves and jointly models fine-grained token and entity signals throughout the ranking process.

    \item \textbf{Bilinear Interaction Modeling:} To effectively combine these rich interaction signals, we replace traditional similarity metrics with bilinear projections, addressing the limitations of dot product and cosine similarity scoring used in prior work.
    \item \textbf{Noise-robust Entity Incorporation:} Entity linking is inherently noisy, but prior work \cite{xiong2017word,liu-etal-2018-entity} shows that attention can mitigate this by emphasizing query-relevant entities. Building on this insight, \texttt{QDER} uses \emph{query-guided attention} to dynamically refine token and entity representations, promoting relevance and suppressing linking errors.

\end{itemize}

\emph{\textbf{This paper makes the following contributions}}:
\begin{itemize}[leftmargin=*]
   \item We propose \texttt{QDER}, a multi-vector neural re-ranking architecture that uses dual text and entity channels with attention-guided interaction modeling to create query-specific document representations. \texttt{QDER} significantly outperforms existing methods on multiple benchmark datasets.
    \item We provide compelling evidence challenging the prevalent use of static embeddings in neural IR. Our experiments show that query-specific embeddings (\texttt{QDER}) dramatically outperform static embeddings (SentenceBERT) in both clustering quality and ranking performance, suggesting a fundamental need to shift away from pre-computed document representations.
    \item We show that Addition and Multiplication operations are nearly statistically independent in capturing query-document relationships and exhibit superior noise robustness compared to Subtraction, offering valuable insights for future architecture design.
    \item We show that entity-attention mechanisms are vital for difficult queries, with \texttt{QDER} achieving an nDCG@20 of 0.70 where traditional methods fail (nDCG@20 = 0.0). This stems from its ability to dynamically prioritize relevant entities across multiple dimensions while filtering out irrelevant ones.
\end{itemize}

\section{Related Work}
\label{sec:Related Work}

\textbf{Entity-Oriented Search.} The field of entity-oriented search has evolved significantly over time. Initially, researchers like \citet{meij2010conceptual} and \citet{dalton2014entity} focused on using entities for query expansion. The field then progressed to treating entities as a latent layer, with notable works including Explicit Semantic Analysis \cite{gabrilovich2009wikipedia} and the Latent Entity Space model \cite{liu2015latent}. EsdRank \cite{xiong2015esdrank} further advanced this approach by creating entity-based connections between queries and documents.
The next wave of research integrated entities as explicit elements in retrieval models. \citet{raviv2016document} and \citet{ensan2017document} developed entity-based language models, while Xiong et al. \cite{xiong2015esdrank,xiong2015query,xiong2016bag,xiong2017end,xiong2017explicit,xiong2017jointsem,xiong2017word,xiong2018towards} pioneered the "bag-of-entities" approach alongside traditional "bag-of-words" methods.

\textbf{Single-Vector Neural IR.} In neural IR, pre-BERT developments fell into two categories: representation-based models \cite{huang2013learning,shen2015entity,mitra2019an,nalisnick2016improving} and interaction-based models \cite{guo2016drmm,xiong2017end,hui-etal-2017-pacrr,hui2018copacrr,dai2018cknrm}. The introduction of BERT \cite{devlin2018bert} and its variants \cite{liu2019roberta,clark2020electra,he2020deberta,zhang-etal-2019-ernie,jiang2020convbert} revolutionized the field, leading to innovations like Birch \cite{akkalyoncu-yilmaz-etal-2019-cross}, BERT-MaxP \cite{dai2019deeper}, CEDR \cite{macavaney2019ceder}, and PARADE \cite{li2020parade}. The Transformer Kernel model \cite{hofstatter2019tu,hofstatter2020interpretable}, later enhanced with local self-attention \cite{hofstatter2020local}, represented another significant architectural innovation.
Bi-encoders emerged as a crucial development, with DPR \cite{karpukhin-etal-2020-dense} and ANCE \cite{xiong2020ance} leading the way. To address bi-encoders' limitations in capturing term-level interactions, ColBERT \cite{khattab2020colbert} introduced ``late interactions,'' while TCT-ColBERT \cite{lin2020distilling} explored knowledge distillation. There is a growing shift toward leveraging embeddings for pseudo-relevance feedback.

CEQE \cite{naseri2021ceqe} selects expansion terms from top-ranked feedback documents that are closest to the query in the BERT embedding space. ColBERT-PRF \cite{xiao2023colbert-prf} uses BERT embeddings for retrieval, thus avoiding topic drifts for polysemous words. 

\textbf{Multi-Vector Neural IR.} Recent neural IR advancements leverage multi-vector representations to capture rich query-document interactions beyond single-vector models. \citet{humeau2020polyencoders} introduced the poly-encoder, which generates multiple document representations using learned “context codes” and aggregates them via an attention mechanism with the query vector for relevance scoring.  \citet{luan-etal-2021-sparse} proposed ME-BERT, where the document encoder produces \(m\) representations from the first \(m\) tokens. The final score is the maximum inner product between the query vector and these \(m\) representations, enabling efficient nearest neighbor search.  \citet{khattab2020colbert} extended this idea with ColBERT, which creates dense token-level representations for documents. Its ``late interaction'' mechanism uses the \texttt{MaxSim} operator to compute relevance by aggregating the maximum cosine similarity between query and document tokens, allowing for fine-grained matching with scalable indexing techniques.

\section{Approach}
\label{sec:Approach} 

The following sections detail our approach: Section \ref{subsec:Conceptual Foundation} outlines the framework's key concepts and guiding hypotheses, while Section \ref{subsec:QDER: Technical Architecture} describes the model architecture, scoring mechanisms, and integration strategies for effective relevance modeling.
\vspace{-0.3cm}

\subsection{QDER: Conceptual Foundation}
\label{subsec:Conceptual Foundation}

At its core, \texttt{QDER} introduces a novel paradigm for document re-ranking, called \textbf{\textit{Attention-Guided Multi-Granular Interaction Decomposition}}. It is based on the hypothesis that document relevance stems from two semantic relationships: direct alignment and complementary context. For example, a query like ``\textit{Steve Jobs' role in the development of the iPhone}'' requires capturing both explicit matches (e.g., ``iPhone'' development) and related context (e.g., Jobs' leadership at Apple), which together shape relevance.

We hypothesize that learned attention patterns can help identify and prioritize semantically relevant aspects of both queries and documents, leading to improved retrieval performance. A core innovation of \texttt{QDER} is its ``late aggregation'' approach: Instead of collapsing token and entity information early, we maintain individual representations throughout the matching process and aggregate them only at final scoring. Unlike prior work \cite{chatterjee2024dreq,reimers2019sentence} that applies interaction operations to static or aggregated embeddings, \texttt{QDER} uses attention-weighted embeddings, allowing the query context to dynamically emphasize the most relevant parts of the document. This attention-guided refinement transforms raw token and entity embeddings into query-specific representations tailored to capture complex semantic relationships.

To effectively model these relationships, \texttt{QDER} employs a dual-channel architecture that captures complementary semantic patterns at different granularities: A \textit{text-based channel} that focuses on fine-grained linguistic alignment and complementarity, and an \textit{entity-aware channel} that models higher-level semantic relationships. Within each channel, attention scores act as dynamic gates, reshaping document representations to emphasize aspects most relevant to the query. Based on our analysis of attention patterns, we observe that in the text channel, higher attention weights tend to correspond with document tokens that appear relevant to query terms. Similarly, in the entity channel, our experiments suggest attention-based adjustments help identify entities that may have semantic relationships with query concepts.

After attention-based matching, the model captures rich interaction patterns using two fundamental operations carefully chosen to model different aspects of relevance: (1) Semantic alignment (via multiplication) identifies precise matches between query and document aspects, and (2) Semantic complementarity (via addition) reveals how document content enriches and expands upon the query intent. As these operations are applied to attention-weighted embeddings, they capture more nuanced relationships than traditional static approaches. For instance, multiplication on attention-weighted representations can identify semantic alignments that are specifically relevant to the query's information need, rather than generic term matches. Similarly, addition operates on embeddings that already emphasize query-relevant aspects, enabling the model to measure complementarity in the context of the query's specific intent. This attention-guided approach to interaction modeling ensures that each operation focuses on aspects that matter most for relevance assessment.
We hypothesize that interactions on query-specific embeddings would enable \texttt{QDER} to better capture alignment and complementarity, essential for understanding document-query relationships. Unlike methods relying on static embeddings, \texttt{QDER} dynamically reshapes embeddings with attention and targeted interactions, ensuring relevance is tailored to each query.

\vspace{-2mm}
\subsection{QDER: Technical Architecture}
\label{subsec:QDER: Technical Architecture}

\subsubsection{Multi-Channel Representation Processing}
\label{subsubsec:Multi-Channel Representation Processing}

To capture both linguistic and semantic signals from documents, \texttt{QDER} processes information through two parallel channels:

\textbf{Text Channel}: The text channel captures fine-grained linguistic patterns through contextualized embeddings:
\begin{itemize}
    \item Query text representation: $Q^t = \text{Encoder}(q) \in \mathbb{R}^{l_q \times d_t}$
    \item Document text representation: $D^t = \text{Encoder}(d) \in \mathbb{R}^{l_d \times d_t}$
\end{itemize}
where $l_q$ and $l_d$ denote sequence lengths, and $d_t$ is the embedding dimension. We use BERT \cite{devlin2018bert} as encoder here.

\textbf{Entity Channel}: Operating in parallel, the entity channel processes higher-level semantic concepts:
\begin{equation}
    Q^e \in \mathbb{R}^{n_q \times d_e} \text{ and } D^e \in \mathbb{R}^{n_d \times d_e}
\end{equation}
where $n_q$, $n_d$ are entity counts and $d_e$ is the embedding dimension. To obtain query entities, we follow prior work \cite{meij2010conceptual}: we pool entities from candidate documents to form a set $\mathcal{E}$. We transfer the relevance labels from documents to their entities, assuming relevant documents contain relevant entities \cite{nguyen2016marco}. We then train a BERT-based entity ranker using entity descriptions from DBpedia to map entities in  $\mathcal{E}$ to queries. Document entities are obtained using entity linking.

\subsubsection{Dynamic Attention-Guided Interaction}
\label{subsubsec:Dynamic Attention-Guided Interaction}
A key innovation in \texttt{QDER} is its attention-first approach to modeling interactions. Traditional methods often compute interactions on query-agnostic (static) representations, potentially diluting important matching signals. Instead, \texttt{QDER} first establishes attention-based matching to identify critical alignments between query and document elements.

Our working hypothesis is that attention weights may correlate with relevance, with higher attention values potentially indicating tokens or entities that are more relevant to the query. Section \ref{subsec:Entity Attention Analysis} provides empirical analysis of this relationship. By aligning query and document elements through learned attention patterns, \texttt{QDER} ensures that subsequent interaction modeling focuses on the most semantically relevant components.

\textbf{Text Channel Attention}: We compute attention weights that highlight relevant document tokens for each query token:
\begin{equation}
    A^t = \text{softmax}(Q^t {D^t}^\top) \in \mathbb{R}^{l_q \times l_d}
\end{equation}
\begin{equation}
    \tilde{D}^t = A^t D^t
\end{equation}

\textbf{Entity Channel Attention}: Similarly for entities:
\begin{equation}
    A^e = \text{softmax}(Q^e {D^e}^\top) \in \mathbb{R}^{n_q \times n_d}
\end{equation}
\begin{equation}
    \tilde{D}^e = A^e D^e
\end{equation}

\subsubsection{Multi-Granular Interaction Modeling}
\label{subsubsec:Multi-Granular Interaction Modeling}
After obtaining attention-weighted representations, \texttt{QDER} models two fundamental types of semantic relationships that together provide a comprehensive view of query-document relevance:

\textbf{1. Semantic Alignment} (via element-wise multiplication):
\begin{equation}
    M^t = Q^t \odot \tilde{D}^t \text{ and } M^e = Q^e \odot \tilde{D}^e
\end{equation}
Multiplication emphasizes areas of strong alignment, highlighting strong semantic similarities. This operation is particularly effective at identifying documents that precisely address specific query aspects, capturing both token-level matches in the text channel and entity-level correspondence in the entity channel.

\textbf{2. Semantic Complementarity} (via addition operations):
\begin{equation}
    C^t = Q^t + \tilde{D}^t \text{ and } C^e = Q^e + \tilde{D}^e
\end{equation}
Addition captures how query and document representations complement each other, revealing overlapping themes and shared contexts. This operation is especially valuable for identifying documents that provide additional relevant context beyond exact matches, helping to model topical relevance and thematic alignment.

Our focus on multiplication and addition operations stems from their complementary roles in capturing relevance signals. Extensive experimentation (Section \ref{subsec:Interaction Operations in Query-Document Matching}) shows that these two operations effectively capture semantic similarity, achieving optimal document ranking performance.

\subsubsection{Relevance Signal Integration}
\label{subsubsec:Relevance Signal Integration}
\texttt{QDER} combines multi-granular interaction patterns with external relevance signals to create a unified and effective scoring mechanism. We aggregate each interaction pattern through mean pooling ($h^t_* = \text{MeanPool}(*)$ for $* \in \{C^t, M^t\}$ and $h^e_* = \text{MeanPool}(*)$ for $* \in \{C^e, M^e\}$), and enhance them with external relevance scores ($\hat{h}^t_* = s \cdot h^t_*$ and $\hat{h}^e_* = s \cdot h^e_*$), where $s$ represents an external score like BM25. Our experiments suggest this approach creates more discriminative query-specific representations. By incorporating traditional retrieval scores as scaling factors, we aim to create query-focused embeddings that potentially better reflect the query's information need. Our experiments in Section \ref{subsec:Query-Specific Embeddings} suggest this approach helps create more discriminative query-specific representations, though the precise mechanism requires further investigation.

\subsubsection{Cross-Pattern Interaction Scoring}
\label{subsubsec:Cross-Pattern Interaction Scoring}

Consider the query: ``\textit{Steve Jobs' role in the development of the iPhone.}'' Documents may show strong entity alignment (Jobs without iPhone), strong text alignment (iPhone without Jobs), or ideally, both. While element-wise operations in \texttt{QDER} can capture individual interactions, they cannot determine how best to combine these signals. For instance, when should broad topical relevance outweigh specific matches? We hypothesize that modeling relationships between interaction patterns enables the model to prioritize balanced evidence and rank documents with complementary text and entity strengths higher.

We propose a two-stage interaction approach distinct from traditional methods that use direct query-document matching (e.g., cosine similarity \cite{karpukhin-etal-2020-dense,reimers2019sentencebertsentenceembeddingsusing}). \texttt{QDER} first computes meaningful interaction patterns through element-wise operations, then learns optimal combinations for ranking.
Given the feature vector concatenating the previously defined alignment and complementarity patterns:
$$h = [h^t_m; h^t_c; h^e_m; h^e_c]$$ where $h^t_m$, $h^e_m$ capture precise matching through multiplication in text and entity space, and $h^t_c$, $h^e_c$ model broader contextual relationships through addition, we compute the final relevance score using a bilinear interaction:
\begin{equation}
\text{score} = h^\top M h = \sum_{i=1}^d \sum_{j=1}^d h_i M_{i,j} h_j
\end{equation}

Here, $M \in \mathbb{R}^{d \times d}$ is a learnable interaction matrix modeling cross-pattern relationships. For example, $M_{1,3}$ might capture how text and entity alignments reinforce each other, while $M_{2,4}$ could learn how complementarity patterns indicate comprehensive coverage. This approach enables \texttt{QDER} to learn sophisticated relevance strategies that adapt to different query types, learning optimal pattern combinations directly from data rather than using predefined rules.

\subsubsection{End-to-End Training}
\label{subsubsec:End-to-End Training}
\texttt{QDER} is trained end-to-end using binary cross-entropy loss:
\begin{equation}
    L = -\frac{1}{N}\sum_{i=1}^N [y_i \log(\hat{y}_i) + (1-y_i)\log(1-\hat{y}_i)]
\end{equation}
where $y_i$ is the ground truth relevance label and $\hat{y}_i$ is the predicted relevance probability. The entire model is optimized end-to-end using back-propagation. This unified training approach allows the model to learn optimal attention patterns and interaction weights simultaneously, ensuring that all components work together effectively to assess document relevance.

\subsubsection{Final Hybrid Scoring}
\label{subsubsec:Final Hybrid Scoring}

To further enhance retrieval performance, following previous work \cite{nogueira2019passage,zhan2020repbertcontextualizedtextembeddings}, we combine the strengths of lexical matching provided by BM25 with the semantic richness of \texttt{QDER} scores through a linear interpolation strategy. 
Given a query $q$ and a document $d$, the final hybrid score is computed as:
\begin{equation*}
    \text{score}_{\text{hybrid}}(q, d) = \lambda \cdot \text{score}_{\text{BM25}}(q, d) + (1 - \lambda) \cdot \text{score}_{\text{QDER}}(q, d)
\end{equation*}
where $\text{score}_{\text{BM25}}(q, d)$ is the relevance score assigned by BM25, $\text{score}_{\text{QDER}}(q, d)$ is the relevance score computed by \texttt{QDER}, and $\lambda \in [0, 1]$ is a parameter that controls the relative contribution of the two components. We learn $\lambda$ using Coordinate Ascent optimized for Mean Average Precision.

\subsubsection{Complementary Integration of Relevance Signals}
\label{subsubsec:Complementary Integration of Relevance Signals}

While \texttt{QDER} uses BM25 scores in both the relevance signal integration (Section \ref{subsubsec:Relevance Signal Integration}) and final hybrid scoring (Section \ref{subsubsec:Final Hybrid Scoring}), these serve distinct roles. In the integration phase, BM25 scores weight interaction patterns, enhancing relevance signals in the learned representations to guide semantic matching. In contrast, the hybrid scoring phase combines \texttt{QDER}'s semantic matching with BM25's exact lexical matching, capturing both semantic and lexical relationships.

\section{Experimental Methodology}
\label{sec:Experimental Methodology}

\subsection{Datasets}
\label{subsec:Datasets}
We evaluate the efficacy of our approach using the following large-scale, established benchmarks for document and passage ranking. Our evaluation spans diverse query sets, two distinct retrieval tasks--document ranking and passage ranking--and two types of collections: news articles and complex answer topics.

\textbf{CODEC.} This benchmark \cite{mackie2022codec} is designed for complex research topics in social sciences. It includes 42 topics, and an entity linked corpus with 729,824 documents with focused content across finance, history, and politics. The corpus contains an average of 159 entities per document. It provides expert judgments on 6,186 documents derived from diverse automatic and manual runs.

\textbf{TREC Complex Answer Retrieval (CAR) 2017.} The dataset~\cite{dietz2017trec} 
is derived from Wikipedia. The queries correspond to article headings and subheadings. In this work, we focus on the \textit{page-level} queries, i.e., the title of the Wikipedia page as the query and address the paragraph ranking task. The documents are paragraphs extracted from Wikipedia articles. We use the \texttt{BenchmarkY1-Train} subset from the TREC CAR v2.1 data release. This subset is based on a Wikipedia dump from 2016. The ground truth is automatic with 117 page-level queries, and 4862 positive passage assessments. 

\textbf{TREC Robust 2004.} The TREC Robust 2004 track \cite{voorhees2003overview} focuses on poorly performing topics. The track provides 250 topics with short ``titles'' and longer ``descriptions''; we report results on both title and description queries. The collection consists of 528,024 documents (containing an average of 116 entities per document) taken from TREC disks 4 and 5 excluding the Congressional Record. The track provides 311,409 graded relevance judgments for evaluation.

\textbf{TREC News 2021.} The TREC News track \cite{soboroff2018trec} focuses on search tasks within the news domain. We focus on the background linking task, which involves retrieving news articles that provide relevant context or background information for a given news story. There are 51 topics, each with a title, description, and narrative; in this work, we use all three fields for query formulation. The track uses the TREC Washington Post (v4) collection, encompassing 728,626 documents containing 131 entities per document on average. The track provides 12,908 graded relevance assessments for evaluation.

\textbf{TREC Core 2018.} The track \cite{allan2017trec} offers 50 topics, each with titles, descriptions, and narratives. For this work, we utilize all three components. It uses the TREC Washington Post (v2) collection, encompassing 595,037 documents with 123 entities per document on average. 26,233 graded relevance judgements are available.

\vspace{-2mm}
\subsection{Evaluation Paradigm}
\label{subsec:Evaluation Paradigm}

\textbf{Candidate Ranking.} We retrieve a candidate set of 1000 documents per query using \texttt{BM25+RM3} (Pyserini default). 
\textbf{Evaluation Metrics.} (1) Precision at $k=20$, (2) Normalized Discounted Cumulative Gain (nDCG) at $k=20$, (3) Mean Average Precision (MAP), and (4) Mean Reciprocal Rank (MRR). 
We use the official \texttt{trec\_eval} tool from NIST (with the \texttt{-c} flag) to evaluate each system.  We conduct significance testing using paired-t-tests. 

\textbf{Entity Embeddings.} We use pre-trained Wikipedia2Vec \cite{yamada-etal-2020-wikipedia2vec} entity embeddings provided by Kamphuis et al. \cite{Kamphuis2023mmead}.

\textbf{Entity Linking.}  We use WAT \cite{piccinno2014wat} entity linker in this work.

\textbf{Train and Test Data.}  As positive examples during training, we use documents that are assessed as relevant in the ground truth provided with the dataset. Following the standard \cite{karpukhin-etal-2020-dense}, for negative examples, we use documents from the candidate ranking (BM25+RM3) which are either explicitly annotated as negative or not present in the ground truth. We balance the training data by keeping the number of negative examples the same as the number of positive examples. These examples are then divided into 5-folds for cross-validation. We create these folds at the query level.

\textbf{Baselines.}
We compare our proposed re-ranking approach \texttt{QDER} to the following supervised state-of-the-art \textbf{neural re-rankers}: (1) \textbf{RoBERTa \cite{liu2019roberta}}, (2) \textbf{DeBERTa \cite{he2020deberta}}, (3) \textbf{ELECTRA \cite{clark2020electra}}, (4) \textbf{ConvBERT \cite{jiang2020convbert}}, (5) \textbf{RankT5 \cite{zhuang2023rankt5}},
(6) \textbf{ERNIE \cite{zhang-etal-2019-ernie}}, (7) \textbf{EDRM \cite{liu-etal-2018-entity}}. 
We include the following \textbf{multi-vector models for re-ranking}: (1) \textbf{ColBERT \cite{khattab2020colbert}}, (2) \textbf{ME-BERT \cite{luan-etal-2021-sparse}}, and (3) \textbf{PolyEncoders \cite{humeau2020polyencoders}}.
On TREC Robust 2004, we include the following additional (full-retrieval) baselines: (1) \textbf{CEDR \cite{macavaney2019ceder}}, (2) \textbf{EQFE \cite{dalton2014entity}}, (3) \textbf{SPLADE \cite{thibault2021splade}}, and 
(4) \textbf{ANCE-MaxP \cite{xiong2020ance}} 
All baselines are fine-tuned on the target datasets via 5-fold cross-validation using the binary cross-entropy.

\subsection{Implementation Details}
\label{subsec:Implementation Details}

We implemented our model using PyTorch and the HuggingFace Transformers library, leveraging the \texttt{bert-base-uncased} model as the base encoder with a maximum sequence length of 512 tokens. The model was fine-tuned using the \texttt{CrossEntropyLoss} function provided by PyTorch. 
For optimization, we employed the Adam optimizer \cite{kingma2014adam} with a learning rate of $2e-5$ and a linear warmup schedule over the first 1,000 steps. The training process used a batch size of 20 and was carried out for 10 epochs, with the model evaluated after each epoch. Early stopping was implemented based on the validation MAP score, calculated using \texttt{pytrec\_eval}, and the best-performing checkpoint was saved.
To enhance inference efficiency, document embeddings and entity links were precomputed and cached offline. All experiments were conducted on a single NVIDIA A6000 Ada GPU with 48GB of memory.

\section{Results and Analysis}
\label{sec:Results and Analysis}

\begin{table*}[t]
\centering
\caption{Overall results TREC Robust 2004. Candidate Ranking (BM25) on the top. Best results in bold. $\blacktriangle$ denotes significant improvement and $\blacktriangledown$ denotes significant deterioration compared to $\star$ using a paired-t-test at $p<0.05$.}
\label{tab:res-tab-1}
\scalebox{0.8}{
\begin{tabular}{
  @{}l
  S[table-format=1.4, table-space-text-post={--}]
  S[table-format=1.4, table-space-text-post={--}]
  S[table-format=1.4, table-space-text-post={--}]
  S[table-format=1.4, table-space-text-post={--}]|
  S[table-format=1.4, table-space-text-post={--}]
  S[table-format=1.4, table-space-text-post={--}]
  S[table-format=1.4, table-space-text-post={--}]
  S[table-format=1.4, table-space-text-post={--}]@{}
}
\toprule
                          & \multicolumn{4}{c}{\textbf{TREC Robust 2004 (Title)}}          & \multicolumn{4}{c}{\textbf{TREC Robust 2004 (Description)}}    \\ \midrule
                          & \textbf{MAP} & \textbf{nDCG@20} & \textbf{P@20} & \textbf{MRR} & \textbf{MAP} & \textbf{nDCG@20} & \textbf{P@20} & \textbf{MRR} \\ \midrule
BM25$\star$              & 0.2915       & 0.4354           & 0.3839        & 0.6693       & 0.2779       & 0.4247           & 0.3681        & 0.6607       \\ \midrule
RankT5 (Enc)             & 0.3028$\blacktriangle$       & 0.4935$\blacktriangle$           & 0.4293$\blacktriangle$        & 0.7455$\blacktriangle$       & 0.3271$\blacktriangle$       & 0.5398$\blacktriangle$           & 0.4558$\blacktriangle$        & 0.8177$\blacktriangle$       \\
BERT                     & 0.2966$\blacktriangle$       & 0.4789$\blacktriangle$           & 0.4094$\blacktriangle$        & 0.7386$\blacktriangle$       & 0.3190$\blacktriangle$       & 0.5232$\blacktriangle$           & 0.4500$\blacktriangle$        & 0.7871$\blacktriangle$       \\
RoBERTa                  & 0.2899$\blacktriangledown$       & 0.4736$\blacktriangle$           & 0.4100$\blacktriangle$        & 0.7310$\blacktriangle$       & 0.3341$\blacktriangle$       & 0.5411$\blacktriangle$           & 0.4602$\blacktriangle$        & 0.8136$\blacktriangle$       \\
DeBERTa                  & 0.2931$\blacktriangle$       & 0.4856$\blacktriangle$           & 0.4217$\blacktriangle$        & 0.7379$\blacktriangle$       & 0.3411$\blacktriangle$       & 0.5465$\blacktriangle$           & 0.4673$\blacktriangle$        & 0.8058$\blacktriangle$       \\
ELECTRA                  & 0.2680$\blacktriangledown$       & 0.4461$\blacktriangle$           & 0.3869$\blacktriangle$        & 0.6882$\blacktriangle$       & 0.2924$\blacktriangle$       & 0.4904$\blacktriangle$           & 0.4141$\blacktriangle$        & 0.7654$\blacktriangle$       \\
ConvBERT                 & 0.3212$\blacktriangle$       & 0.5185$\blacktriangle$           & 0.4506$\blacktriangle$        & 0.7648$\blacktriangle$       & 0.3498$\blacktriangle$       & 0.5665$\blacktriangle$           & 0.4839$\blacktriangle$        & 0.8317$\blacktriangle$       \\ \midrule
ColBERT v2               & 0.2915       & 0.4730$\blacktriangle$           & 0.4102$\blacktriangle$        & 0.7095$\blacktriangle$       & 0.2880$\blacktriangle$       & 0.4806$\blacktriangle$           & 0.4120$\blacktriangle$        & 0.7446$\blacktriangle$       \\
ME-BERT                  & 0.0774$\blacktriangledown$       & 0.0946$\blacktriangledown$           & 0.0976$\blacktriangledown$        & 0.1660$\blacktriangledown$       & 0.0575$\blacktriangledown$       & 0.0749$\blacktriangledown$           & 0.0733$\blacktriangledown$        & 0.1476$\blacktriangledown$       \\
PolyEncoder              & 0.0753$\blacktriangledown$       & 0.0984$\blacktriangledown$           & 0.0974$\blacktriangledown$        & 0.2003$\blacktriangledown$       & 0.0623$\blacktriangledown$       & 0.0820$\blacktriangledown$           & 0.0825$\blacktriangledown$        & 0.1858$\blacktriangledown$       \\ \midrule
BERT-MaxP                & 0.3174$\blacktriangle$       & 0.4853$\blacktriangle$           & 0.4205$\blacktriangle$        & 0.7331$\blacktriangle$       & 0.3149$\blacktriangle$       & 0.4916$\blacktriangle$           & 0.4175$\blacktriangle$        & 0.7697$\blacktriangle$       \\
CEDR                     & 0.3701$\blacktriangle$       & 0.5475$\blacktriangle$           & 0.4769$\blacktriangle$        & 0.7879$\blacktriangle$       & 0.4000$\blacktriangle$       & 0.5983$\blacktriangle$           & 0.5185$\blacktriangle$        & 0.8457$\blacktriangle$       \\
SPLADE                   & 0.2243$\blacktriangledown$       & 0.4202$\blacktriangledown$           & 0.3588$\blacktriangledown$        & 0.6827$\blacktriangle$       & 0.2314$\blacktriangledown$       & 0.4309$\blacktriangle$           & 0.3600$\blacktriangledown$        & 0.7319$\blacktriangle$       \\
ANCE-MaxP                & 0.1325$\blacktriangledown$       & 0.3087$\blacktriangledown$           & 0.2496$\blacktriangledown$        & 0.6095$\blacktriangledown$       & & & & \\ \midrule
ERNIE                    & 0.2894$\blacktriangledown$       & 0.4753$\blacktriangle$           & 0.4116$\blacktriangle$        & 0.7135$\blacktriangle$       & 0.3278$\blacktriangle$       & 0.5373$\blacktriangle$           & 0.4546$\blacktriangle$        & 0.8058$\blacktriangle$       \\
EDRM                     & 0.0671$\blacktriangledown$       & 0.1028$\blacktriangledown$           & 0.0940$\blacktriangledown$        & 0.2413$\blacktriangledown$       & 0.0486$\blacktriangledown$       & 0.0687$\blacktriangledown$           & 0.0679$\blacktriangledown$        & 0.1721$\blacktriangledown$       \\
EQFE                     & 0.3278$\blacktriangle$       & 0.4307$\blacktriangledown$           & 0.3797$\blacktriangledown$        & 0.6499$\blacktriangledown$       & & & & \\
ExactMatch               & 0.2030$\blacktriangledown$       & 0.4759$\blacktriangle$           & 0.3988$\blacktriangle$        & 0.9152$\blacktriangle$       & 0.2029$\blacktriangledown$       & 0.4617$\blacktriangle$           & 0.3813$\blacktriangle$        & 0.4617$\blacktriangledown$       \\ \midrule
TREC Best                & 0.3338$\blacktriangle$       &                & 0.4275$\blacktriangle$        &            & 0.3338$\blacktriangle$       &                & 0.4275$\blacktriangle$        &            \\ \midrule
\textbf{QDER}            & $\mathbf{0.6082}$$\blacktriangle$       & $\mathbf{0.7694}$$\blacktriangle$           & $\mathbf{0.7361}$$\blacktriangle$        & $\mathbf{0.9751}$$\blacktriangle$       & $\mathbf{0.5855}$$\blacktriangle$       & $\mathbf{0.7516}$$\blacktriangle$           & $\mathbf{0.7120}$$\blacktriangle$        & $\mathbf{0.9727}$$\blacktriangle$       \\
\bottomrule
\end{tabular}
}
\end{table*}

\begin{table*}[t]
\centering
\caption{Overall results on TREC News 2021 and TREC Core 2018. Only best performing baselines shown.}
\label{tab:res-tab-2}
\scalebox{0.8}{
\begin{tabular}{
  @{}l
  S[table-format=1.4]
  S[table-format=1.4]
  S[table-format=1.4]
  S[table-format=1.4]|
  S[table-format=1.4]
  S[table-format=1.4]
  S[table-format=1.4]
  S[table-format=1.4]@{}
}
\toprule
                          & \multicolumn{4}{c}{\textbf{TREC Core 2018}}                    & \multicolumn{4}{c}{\textbf{TREC News 2021}}                    \\ \midrule
                          & \textbf{MAP} & \textbf{nDCG@20} & \textbf{P@20} & \textbf{MRR} & \textbf{MAP} & \textbf{nDCG@20} & \textbf{P@20} & \textbf{MRR} \\ \midrule
BM25$\star$              & 0.3151       & 0.4470            & 0.4590         & 0.6518       & 0.4680        & 0.4785           & 0.5775        & 0.7767       \\ \midrule
RankT5 (Enc)             & 0.2163$\blacktriangledown$       & 0.3255$\blacktriangledown$           & 0.3340$\blacktriangledown$         & 0.5182$\blacktriangledown$       & 0.2712$\blacktriangledown$       & 0.3124$\blacktriangledown$           & 0.3608$\blacktriangledown$        & 0.6288$\blacktriangledown$       \\
BERT                     & 0.3024$\blacktriangledown$       & 0.4686$\blacktriangle$           & 0.4610$\blacktriangle$         & 0.7271$\blacktriangle$       & 0.4611$\blacktriangledown$       & 0.5112$\blacktriangle$           & 0.5784$\blacktriangle$        & 0.8518$\blacktriangle$       \\
RoBERTa                  & 0.2596$\blacktriangledown$       & 0.3612$\blacktriangledown$           & 0.3880$\blacktriangledown$         & 0.4969$\blacktriangledown$       & 0.4721$\blacktriangle$       & 0.5215$\blacktriangle$           & 0.5843$\blacktriangle$        & 0.8971$\blacktriangle$       \\
DeBERTa                  & 0.3461$\blacktriangle$       & 0.5194$\blacktriangle$           & 0.5070$\blacktriangle$         & 0.7385$\blacktriangle$       & 0.4318$\blacktriangledown$       & 0.4742$\blacktriangledown$           & 0.5578$\blacktriangledown$        & 0.7867$\blacktriangle$       \\
ELECTRA                  & 0.2400$\blacktriangledown$       & 0.3531$\blacktriangledown$           & 0.3610$\blacktriangledown$         & 0.4962$\blacktriangledown$       & 0.4129$\blacktriangledown$       & 0.4619$\blacktriangledown$           & 0.5294$\blacktriangledown$        & 0.8486$\blacktriangle$       \\
ConvBERT                 & 0.3214$\blacktriangle$       & 0.4961$\blacktriangle$           & 0.4890$\blacktriangle$         & 0.7599$\blacktriangle$       & 0.4441$\blacktriangledown$       & 0.4738$\blacktriangledown$           & 0.5500$\blacktriangledown$        & 0.7872$\blacktriangle$       \\ \midrule
ColBERT v2               & 0.2669$\blacktriangledown$       & 0.4552$\blacktriangle$           & 0.4510$\blacktriangledown$         & 0.6760$\blacktriangle$       & 0.4039$\blacktriangledown$       & 0.4624$\blacktriangledown$           & 0.5353$\blacktriangledown$        & 0.7931$\blacktriangle$       \\ \midrule
ERNIE                    & 0.3401$\blacktriangle$       & 0.5195$\blacktriangle$           & 0.5070$\blacktriangle$         & 0.7704$\blacktriangle$       & 0.4808$\blacktriangle$       & 0.5284$\blacktriangle$           & 0.6000$\blacktriangle$        & 0.8736$\blacktriangle$       \\
EDRM                     & 0.0924$\blacktriangledown$       & 0.0984$\blacktriangledown$           & 0.1343$\blacktriangledown$         & 0.3283$\blacktriangledown$       & 0.0924$\blacktriangledown$       & 0.0984$\blacktriangledown$           & 0.1343$\blacktriangledown$        & 0.3283$\blacktriangledown$       \\
ExactMatch               & 0.1540$\blacktriangledown$       & 0.3482$\blacktriangledown$           & 0.3400$\blacktriangledown$         & 0.7273$\blacktriangle$       & 0.1827$\blacktriangledown$       & 0.2793$\blacktriangledown$           & 0.3422$\blacktriangledown$        & 0.8879$\blacktriangle$       \\ \midrule
TREC Best                & 0.4303$\blacktriangle$       &                & 0.6111$\blacktriangle$         &            & 0.4319$\blacktriangledown$       & 0.4688$\blacktriangledown$           & 0.5902$\blacktriangle$        &            \\ \midrule
\textbf{QDER}            & $\mathbf{0.5449}$$\blacktriangle$       & $\mathbf{0.6562}$$\blacktriangle$           & $\mathbf{0.689}$$\blacktriangle$         & $\mathbf{0.8824}$$\blacktriangle$       & $\mathbf{0.6821}$$\blacktriangle$       & $\mathbf{0.5802}$$\blacktriangle$           & $\mathbf{0.748}$$\blacktriangle$        & $\mathbf{0.9299}$$\blacktriangle$       \\
\bottomrule
\end{tabular}
}
\end{table*}

\begin{table*}[t]
\centering
\caption{Overall results on TREC CAR and CODEC. Only best performing baselines shown.}
\label{tab:res-tab-3}
\scalebox{0.8}{
\begin{tabular}{
  @{}l
  S[table-format=1.4]
  S[table-format=1.4]
  S[table-format=1.4]
  S[table-format=1.4]|
  S[table-format=1.4]
  S[table-format=1.4]
  S[table-format=1.4]
  S[table-format=1.4]@{}
}
\toprule
                          & \multicolumn{4}{c}{\textbf{TREC CAR Y1Train (Page-Level)}}                    & \multicolumn{4}{c}{\textbf{CODEC}} \\ \midrule
                          & \textbf{MAP} & \textbf{nDCG@20} & \textbf{P@20} & \textbf{MRR} & \textbf{MAP} & \textbf{nDCG@20} & \textbf{P@20} & \textbf{MRR} \\ \midrule
BM25$\star$              & 0.1586       & 0.2812           & 0.2415        & 0.5637       & 0.3627       & 0.3795           & 0.4321        & 0.6774       \\ \midrule
RankT5 (Enc)             & 0.2225$\blacktriangle$       & 0.5128$\blacktriangle$           & 0.4013$\blacktriangle$        & 0.8895$\blacktriangle$       & 0.3804$\blacktriangle$       & 0.4374$\blacktriangle$           & 0.4524$\blacktriangle$        & 0.8154$\blacktriangle$       \\
BERT                     & 0.2919$\blacktriangle$       & 0.4863$\blacktriangle$           & 0.4120$\blacktriangle$        & 0.8117$\blacktriangle$       & 0.3816$\blacktriangle$       & 0.4400$\blacktriangle$           & 0.4631$\blacktriangle$        & 0.8163$\blacktriangle$       \\
RoBERTa                  & 0.2919$\blacktriangle$       & 0.4872$\blacktriangle$           & 0.4115$\blacktriangle$        & 0.8119$\blacktriangle$       & 0.3594$\blacktriangledown$       & 0.3773$\blacktriangledown$           & 0.4333$\blacktriangle$        & 0.7043$\blacktriangle$       \\
DeBERTa                  & 0.3063$\blacktriangle$       & 0.5125$\blacktriangle$           & 0.4299$\blacktriangle$        & 0.8411$\blacktriangle$       & 0.3897$\blacktriangle$       & 0.4488$\blacktriangle$           & 0.4631$\blacktriangle$        & 0.7698$\blacktriangle$       \\
ELECTRA                  & 0.2617$\blacktriangle$       & 0.4474$\blacktriangle$           & 0.3705$\blacktriangle$        & 0.7848$\blacktriangle$       & 0.3232$\blacktriangledown$       & 0.3174$\blacktriangledown$           & 0.3845$\blacktriangledown$        & 0.5878$\blacktriangledown$       \\
ConvBERT                 & 0.3173$\blacktriangle$       & 0.5203$\blacktriangle$           & 0.4355$\blacktriangle$        & 0.8597$\blacktriangle$       & 0.3816$\blacktriangle$       & 0.4407$\blacktriangle$           & 0.4619$\blacktriangle$        & 0.7670$\blacktriangle$       \\ \midrule
ColBERT v2               & 0.2745$\blacktriangle$       & 0.4538$\blacktriangle$           & 0.3859$\blacktriangle$        & 0.7962$\blacktriangle$       & 0.3745$\blacktriangle$       & 0.4460$\blacktriangle$           & 0.4560$\blacktriangle$        & 0.7921$\blacktriangle$       \\ \midrule
ERNIE                    & 0.3017$\blacktriangle$       & 0.4967$\blacktriangle$           & 0.4201$\blacktriangle$        & 0.7697$\blacktriangle$       & 0.3906$\blacktriangle$       & 0.4574$\blacktriangle$           & 0.4738$\blacktriangle$        & 0.7952$\blacktriangle$       \\
EDRM                     & 0.0176$\blacktriangledown$       & 0.0182$\blacktriangledown$           & 0.0184$\blacktriangledown$        & 0.0634$\blacktriangledown$       & 0.2981$\blacktriangledown$       & 0.2892$\blacktriangledown$           & 0.3524$\blacktriangledown$        & 0.6211$\blacktriangledown$       \\
ExactMatch               & 0.1463$\blacktriangledown$       & 0.3903$\blacktriangle$           & 0.2816$\blacktriangle$        & 0.8002$\blacktriangle$       & 0.2674$\blacktriangledown$       & 0.4120$\blacktriangle$           & 0.4619$\blacktriangle$        & 0.8968$\blacktriangle$       \\ \midrule
\textbf{QDER}            & $\mathbf{0.3638}$$\blacktriangle$ & $\mathbf{0.5819}$$\blacktriangle$ & $\mathbf{0.4940}$$\blacktriangle$ & $\mathbf{0.8768}$$\blacktriangle$ & $\mathbf{0.5210}$$\blacktriangle$ & $\mathbf{0.5035}$$\blacktriangle$ & $\mathbf{0.5810}$$\blacktriangle$ & $\mathbf{0.9522}$$\blacktriangle$ \\
\bottomrule
\end{tabular}
}
\end{table*}

In this section, we present the results of comprehensive experiments evaluating the effectiveness of \texttt{QDER} in document re-ranking. We begin by summarizing the overall performance of \texttt{QDER} on the datasets described in Section \ref{subsec:Datasets}. Detailed experiments exploring specific aspects of \texttt{QDER}'s performance are then presented in Sections \ref{subsec:Overall Results} to \ref{subsec:Architectural Choices}. \textbf{Unless otherwise specified, all analyses is conducted on the TREC Robust 2004 dataset using title queries due to lack of space.} 
\subsection{Overall Results}
\label{subsec:Overall Results}

From Tables \ref{tab:res-tab-1} through \ref{tab:res-tab-3}, we observe that \texttt{QDER} outperforms all baseline models across all evaluation metrics and datasets. For instance, on the TREC Robust 2004 collection with title queries, \texttt{QDER} achieves an $\text{nDCG@20}$ of 0.75, representing a significant 36\% improvement over the strongest baseline, \texttt{CEDR} ($\text{nDCG@20}=0.55$). To contextualize this advancement, \texttt{CEDR} improved the nDCG@20 of the initial \texttt{BM25+RM3} candidate set by 25\% (0.44 to 0.55). In contrast, \texttt{QDER} achieves a remarkable 70\% improvement (0.44 to 0.75).

\begin{figure}[t]
    \centering
    \includegraphics [scale=0.3]{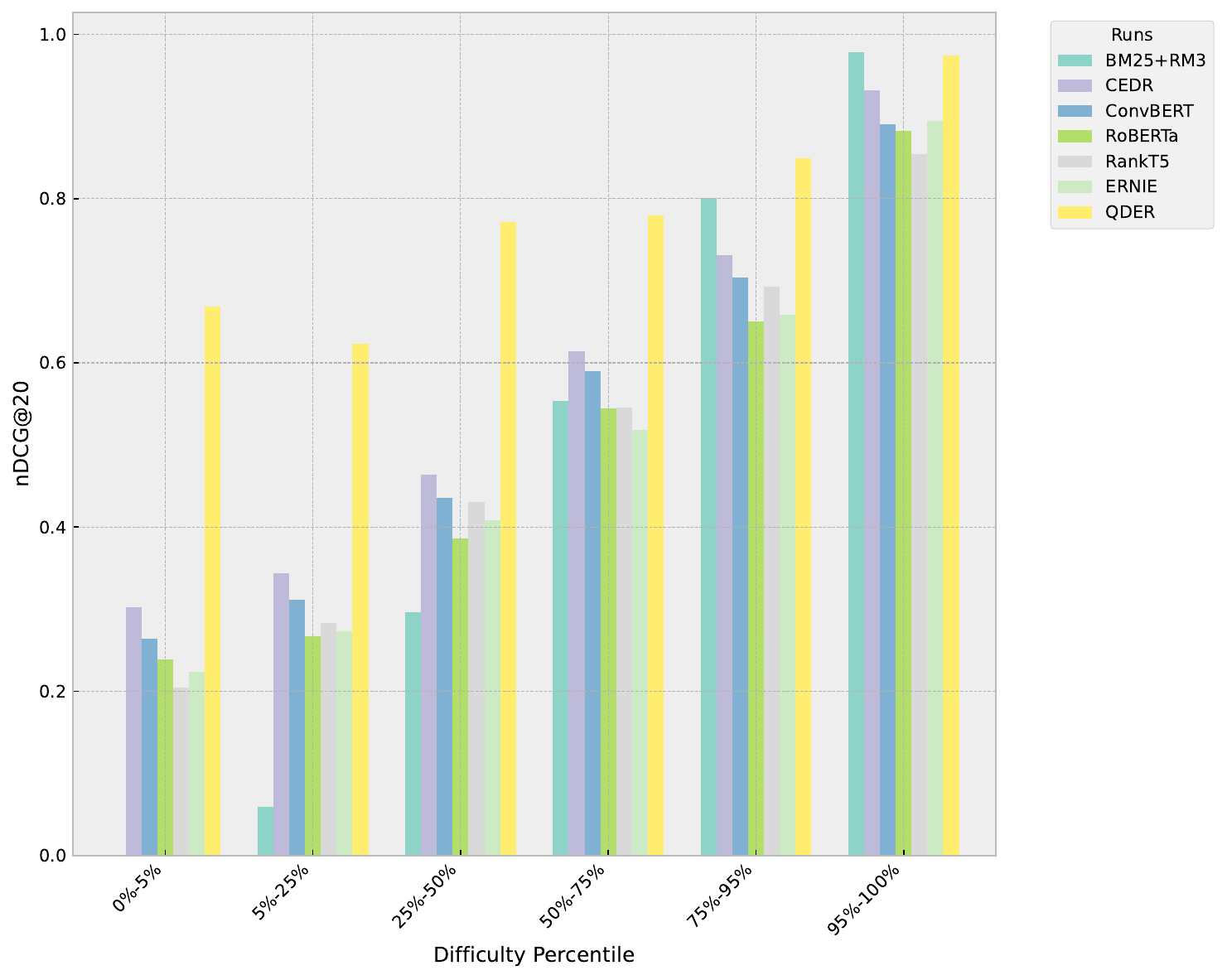}
    \caption{Difficulty test on Robust04 title queries. 5\% most difficult queries for \texttt{BM25+RM3} to the left and the 5\% easiest ones to the right. Performance reported as macro-averages across queries.  For the most difficult queries (0-5\%), relevant documents are promoted to the top of the ranking by \texttt{QDER}.
    }
    \label{fig:diff-test}
\end{figure}

\textbf{Query-Level Analysis.} To further explore these substantial gains, we conduct a fine-grained query-level analysis. By stratifying queries based on their difficulty (measured by \texttt{BM25+RM3} performance), we uncovered a particularly notable insight: \texttt{QDER} excels in addressing the most challenging queries, where improvements are most critical. This is illustrated in Figure \ref{fig:diff-test}. The most pronounced improvements are observed in the bottom quartile (0–25\%), with the most dramatic gains in the extremely challenging 0–5\% bin. For these queries, where traditional methods fail to place any relevant documents in the top-20 positions ($\text{nDCG@20}=0.0$), \texttt{QDER} achieves an impressive $\text{nDCG@20}$ of 0.70, doubling \texttt{CEDR}’s performance ($\text{nDCG@20}=0.35$). \texttt{QDER} improves performance for 211 of 250 queries, compared to 160 by \texttt{CEDR}, demonstrating its robustness across query difficulty levels. Notably, it excels on both ``easy'' queries and the most ``difficult'' ones, making it highly effective for real-world applications with diverse query challenges.

\textbf{Case Study.} A case study further highlights \texttt{QDER}’s effectiveness. Consider the query ``Transportation Tunnel Disasters,'' an example from the most challenging 0–5\% bin. Notably, NIST has classified this query as a Category 5 difficult topic, defined as one requiring multiple aspects that systems often handle inconsistently ~\cite{voorhees2003overview}. \texttt{QDER} demonstrates exceptional capability in surfacing relevant content for this difficult query: It promotes a highly relevant document (as assessed by NIST) from position 734 in the \texttt{BM25+RM3} ranking to position 6, significantly outperforming \texttt{CEDR}’s improvement to position 85. We will revisit this case and analyze in greater depth the mechanisms behind this performance later in Section \ref{subsec:Entity Attention Analysis}.

\subsection{Role of Interaction Operations}
\label{subsec:Interaction Operations in Query-Document Matching}

The effectiveness of \texttt{QDER} fundamentally depends on how its interaction operations transform document representations in response to query context. Our analysis, grounded in comprehensive ablation studies and stability experiments, reveals key insights into why certain operations outperform others, providing both empirical evidence and theoretical validation for their roles.

\begin{table}[t]
\centering
\caption{Ablation Study for Interaction Operations in \texttt{QDER}. Results reported on TREC Robust 2004 (Title).}
\label{tab:interaction-ablation-study}
\scalebox{0.8}{
\begin{tabular}{lcccc}
\toprule
\textbf{Configuration}         & \textbf{MAP}  & \textbf{nDCG@20} & \textbf{P@20} & \textbf{MRR}  \\
\midrule
No-Add                  & 0.5289        & 0.7347           & 0.7008        & 0.9630        \\
No-Multiply             & 0.5492        & 0.7433           & 0.7090        & 0.9594        \\ \midrule
Only-Add                & 0.4703        & 0.6682           & 0.6289        & 0.9104        \\
Only-Subtract           & 0.5230        & 0.7202           & 0.6855        & 0.9559        \\
Only-Multiply           & 0.5325        & 0.7275           & 0.6892        & 0.9726        \\ \midrule
No-Interactions         & 0.2593        & 0.3420           & 0.5011        & 0.3239        \\ \midrule
All-Interactions &0.5091	& 0.7176	& 0.6807	& 0.9598 \\ \midrule
No-Subtract (Ours)            & 0.5667        & 0.7610           & 0.7313        & 0.9714        \\
\bottomrule
\end{tabular}
}
\end{table}

\textbf{Overall Results.} Ablation studies (Table~\ref{tab:interaction-ablation-study}) show that using only Addition and Multiplication operations (\texttt{No-Subtract}) yields the best performance ($\text{nDCG@20} = 0.76$), surpassing \texttt{No-Add} ($0.73$), \texttt{No-Multiply} ($0.74$), and single-operation variants. Removing all operations drastically reduces performance ($\text{nDCG@20} = 0.34$), confirming their importance for query-document matching.

\textbf{Correlation Analysis.} The superiority of the Add-Multiply combination motivates a deeper investigation into the unique roles these operations play. Our correlation analysis reveals that Addition and Multiplication exhibit near-zero (Spearman's) correlation ($-0.022$), suggesting they capture orthogonal and largely independent aspects of query-document relationships.

Addition combines query and document features directly, effectively identifying exact matches between query terms and document terms. For example, when a user searches for   ``\textit{python programming}'', Addition ensures that documents explicitly containing these terms are prioritized. Multiplication, on the other hand, serves as an importance-weighting mechanism, capturing interaction effects between terms. In a query like ``\textit{python web development}'', a document containing ``Django web framework'' might be more relevant than one containing just ``python'' and ``web'' separately, even though both contain the query terms. Together, these operations allow \texttt{QDER} to balance the identification of relevant terms with the nuanced understanding of their interactions.

\textbf{Noise Sensitivity Analysis.} While the Subtraction operation might initially appear complementary, with weak negative correlation to Addition ($-0.218$) and negligible correlation to Multiplication ($0.011$), our analysis reveals its redundancy. We conducted a systematic noise sensitivity analysis by injecting controlled Gaussian noise ($\sigma$ ranging from $0.001$ to $0.1$) into the input embeddings. We measured three key stability metrics: (1) Angular deviation: How much the embeddings' directions changed (in degrees) when noise was added, (2) Noise amplification ratio: How much the operation magnified the input noise, calculated as the ratio of output noise to input noise, and (3) Ranking stability: How well the relative ordering of documents was preserved under noise, measured using Kendall's $\tau$ correlation. 

This analysis reveals stark contrasts in operational stability. Addition and Multiplication demonstrate remarkable robustness, with minimal angular deviations ($2.55\degree$ and $1.66\degree$ respectively) and controlled noise amplification ratios ($1.85$ and $1.89$). Subtraction, however, exhibits pathological behavior with a $9.97\degree$ mean angular deviation—nearly six times higher than Multiplication--and an excessive noise amplification ratio of $5.56$. Although Subtraction achieves the highest ranking stability ($\tau = 0.87$) compared to Addition ($\tau = 0.63$) and Multiplication ($\tau = 0.77$), this result must be interpreted alongside its instability. The high $\tau$ stems from Subtraction consistently exaggerating existing patterns rather than preserving meaningful and robust feature relationships, making its stability counterproductive to effective query-document matching.
\vspace{-1mm}
\subsection{Query-Specific Embeddings}
\label{subsec:Query-Specific Embeddings}

To investigate the impact of attention-guided interactions on document representation quality, we compare query-dependent embeddings produced by \texttt{QDER} with static document embeddings produced by \texttt{SentenceBERT} \cite{reimers2019sentence} (\texttt{SBERT}), a widely used embedding model known for its strong performance in IR. 
Our evaluation focuses on two key aspects: clustering and ranking performance.

\textbf{Clustering Analysis.} First, we examine relevance-based clustering, which groups documents according to their relevance to specific queries. The quantitative metrics from this analysis (Figure 2) show \texttt{QDER} achieves significantly better Davies-Bouldin Index (DBI) (3.36 vs. 24.99) and Silhouette Coefficient (SC) (0.31 vs. 0.002) compared to \texttt{SBERT}, indicating tighter and better-separated relevance clusters.
Second, we analyze topic-wise clustering, where documents are grouped based on their NIST topic assignments. Here, we observe an intriguing pattern that validates our query-specific approach: while \texttt{QDER} shows higher cluster overlap (reflected in topic-wise DBI of 9.19 vs 4.94 and SC of -0.09 vs 0.03), it achieves substantially stronger topic separation as measured by the Calinski-Harabasz Index (201.10 vs 51.42). This seemingly contradictory behavior aligns with \texttt{QDER}'s design---documents relevant to multiple topics adapt their representations based on the query context. This dynamic adjustment allows \texttt{QDER} to capture more nuanced and effective relationships between documents and queries.

These analyses demonstrate the power of \texttt{QDER}'s query-aware document representations: while maintaining clear separation between relevant and non-relevant documents for specific queries, the model also captures the multi-faceted nature of document-topic relationships, ultimately enhancing retrieval quality.

\textbf{Rank Position Analysis.} A detailed evaluation of ranking performance provides compelling evidence of \texttt{QDER}'s superiority over \texttt{SBERT} in surfacing relevant content. The improvements are particularly striking in top-rank positions: \texttt{QDER} more than doubles the number of relevant documents in the top-10 positions (99 versus 46), while concentrating the majority of relevant documents within the top-50 positions. This is further shown by only 20 relevant documents remaining beyond rank 100, highlighting \texttt{QDER}'s effectiveness in elevating relevant content.
The analysis becomes even more revealing when examining performance across NIST relevance grades. For documents rated as highly relevant (NIST grade 2), \texttt{QDER} achieves substantial improvements, raising their average rank from 588.52 (\texttt{BM25+RM3}) to 138.48 and successfully positioning 13 of 21 such documents within the top-50 positions. The impact is even more pronounced for documents rated as relevant (NIST grade 1), where \texttt{QDER} dramatically improves the average rank from 686.78 to 31.47. In this category, \texttt{QDER} demonstrates remarkable precision, promoting 95 documents to top-10 positions and an additional 84 to top-50 positions, achieving a 78.5\% improvement rate.

\textbf{Take-Away.} These findings challenge the prevalent practice of using pre-computed document embeddings in neural IR and suggests a critical architectural shift: future IR systems may need to move beyond static representations to dynamically adapted, query-specific document representations. While this shift introduces new computational challenges, our results demonstrate that the potential gains in ranking quality make this a promising direction for advancing the state-of-the-art in neural IR. \texttt{QDER}'s ability to prioritize relevant documents is particularly valuable for RAG systems, which often rely on the top retrieved documents. By concentrating relevant content in top positions, \texttt{QDER} can significantly enhance the accuracy and context of generated responses.

\begin{figure*}[t]
    \centering
    \includegraphics [scale=0.4]{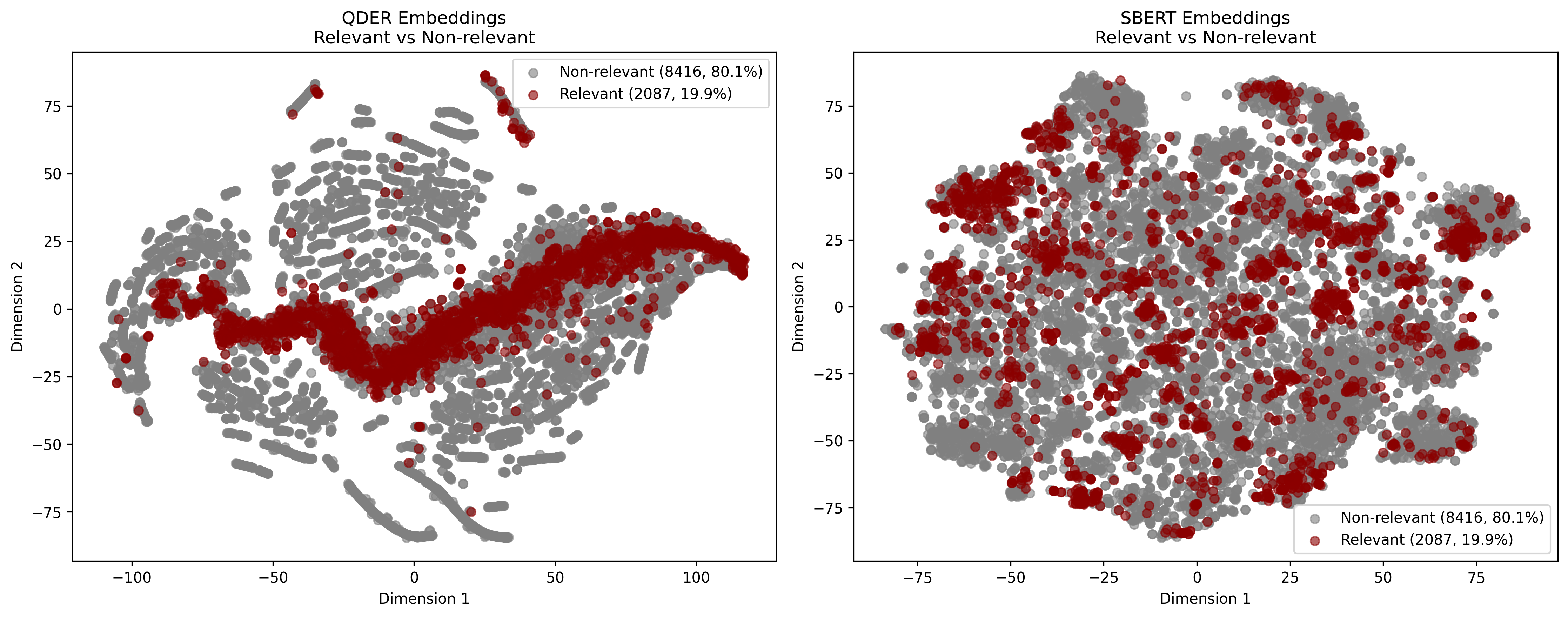}
    \caption{ Visualization of document embeddings using t-SNE for QDER (left) and SBERT (right) on TREC Robust 2004 (Title). Relevant documents (red, 19.8\%) and non-relevant documents (gray, 80.1\%) are shown. QDER produces clear, coherent clusters of relevant documents, while SBERT embeddings display significant mixing. This highlights QDER’s ability to create query-specific representations that naturally separate by relevance, supporting its superior retrieval performance.
    }
    \label{fig:relevance-clustering}
\end{figure*}

\vspace{-1mm}
\subsection{Entity Attention Analysis}
\label{subsec:Entity Attention Analysis}

As discussed in Section \ref{subsec:Conceptual Foundation}, the entity channel in \texttt{QDER} leverages attention-based adjustments to emphasize entities that align with or complement the query's concepts. To illustrate, we revisit the query ``\textit{Transportation Tunnel Disasters}'' from Section \ref{subsec:Overall Results} and analyze how these mechanisms drive \texttt{QDER}’s superior performance.

A closer examination of the entity attention patterns reveals the mechanisms driving this improvement. The model's entity-aware attention identifies and prioritizes relevant entities across multiple dimensions, creating a nuanced understanding of relevance. Safety-critical entities like ``blind curve'' receive strong attention scores (0.33--0.55), highlighting their importance to the query. Location entities such as ``Mission Hills, Los Angeles'' anchor the query to specific incidents with maximum attention scores (1.0). Additionally, transportation-related entities (e.g., ``Sprinter,'' ``101 Freeway,'' ``tunnel junctions'') and safety features (e.g., ``oxygen system'') exhibit coherent attention patterns that align with the query's intent. Notably, the mechanism's strong discriminative power ensures that relevant entities are prioritized, while less pertinent entities receive minimal attention despite frequent appearances in the document.

Our analysis of attention patterns suggests the entity-aware attention mechanism may serve as a contextual filter that helps prioritize certain document sections. For example, we observed high attention scores for entities like ``Mission Hills, Los Angeles'' in documents containing specific incident descriptions related to the query This filtering effect is particularly valuable for long documents (such as in Robust04) where traditional models might struggle to identify the most relevant passages. Moreover, the entity-aware attention mechanism helps the model build a more comprehensive understanding of document relevance by capturing meaningful entity relationships across different aspects of relevance (infrastructure, safety, location). This multi-faceted enhancement of the relevance matching process explains \texttt{QDER}'s ability to dramatically improve rankings for challenging queries. The entity attention patterns effectively guide the model's other components, enabling it to identify and appropriately weight relevant documents even when surface-level matching signals are weak.

This analysis underscores a fundamental difference in the use of entity attention between \texttt{QDER} and prior work \cite{liu-etal-2018-entity,xiong2017word}. In \texttt{EDRM} \cite{liu-etal-2018-entity}, for example, ``attention'' is based on static, precomputed similarities between query and document entities. In contrast, \texttt{QDER} introduces query-specific entity attention that dynamically learns and adjusts entity weights during training, optimizing their relevance to the specific query. By delaying aggregation and maintaining separate entity representations throughout the pipeline, \texttt{QDER} enables truly dynamic entity attention that adapts to each query's specific information needs. For example, \texttt{QDER} can adjust the importance of entities like ``blind curve,'' and ``Mission Hills'' specifically for a transportation disasters query, achieving nuanced, query-aware reasoning beyond the capabilities of static scoring.
\subsection{Architectural Choices}
\label{subsec:Architectural Choices}

\begin{table}[t]
    \centering
    \caption{Ablation Study for Architectural Choices in \texttt{QDER}. Results reported on TREC Robust 2004 (Title).}
    \scalebox{0.8}{
    \begin{tabular}{lcccc}
        \toprule
        \textbf{Model Variant} & \textbf{MAP} & \textbf{nDCG@20} & \textbf{P@20} & \textbf{MRR} \\
        \midrule
        Relevance Signal Integration: None & 0.4286 & 0.5748 & 0.5586 & 0.7791 \\
        Score Method: Linear & 0.3222 & 0.5022 & 0.4496 & 0.8063 \\
        Entity Importance: Entity Ranking & 0.4543 & 0.6554 & 0.6215 & 0.9071 \\
        No Entities & 0.2098 & 0.3519 & 0.3042 & 0.6011 \\
        \bottomrule
    \end{tabular}
    }
    \label{tab:architecture-ablation-study}
\end{table}

As a final analysis, we conduct ablation studies to validate \texttt{QDER}'s key architectural components. 
The results reveal three critical insights. First, bilinear scoring is essential---on Robust04 (title), the linear variant achieves $\text{MAP}=0.32$ compared to \texttt{QDER}'s 0.56, demonstrating that simple linear combinations cannot capture complex query-document relationships. Second, entity modeling proves crucial--removing entities causes MAP to drop to 0.20, while replacing entity attention with direct entity ranking ($\texttt{MAP}=0.45$) still significantly underperforms the full architecture. Our ablation studies suggest that entity attention mechanisms, which adjust entity representations based on query context, contribute significantly to model performance compared to direct entity ranking alone. This may indicate that dynamic entity representations capture more nuanced relationships between queries and documents.

These findings collectively validate \texttt{QDER}'s architectural decisions. The consistent performance improvements across all metrics demonstrate that our approach successfully balances multiple forms of evidence when determining document relevance. While each component contributes meaningfully, it is their careful integration that enables \texttt{QDER} to achieve superior ranking effectiveness.


\vspace{-2mm}
\section{Conclusion}
\label{sec:Conclusion}

We present \texttt{QDER}, a state-of-the-art neural document re-ranking model that leverages attention-guided interaction modeling to create query-specific representations. Our results definitively answer the question (raised in Section \ref{sec:Introduction}) of whether integrating entities into multi-vector frameworks can advance IR performance: \texttt{QDER}'s substantial improvements over both entity-based and multi-vector baselines demonstrate that unifying these approaches yields powerful synergies. Our comprehensive analysis reveals several key insights: (1) performing semantic matching on attention-weighted embeddings rather than raw representations leads to dramatic improvements in ranking quality, (2) Addition and Multiplication operations exhibit mathematical complementarity in capturing different aspects of relevance, and (3) dynamically adapting document representations to query context is crucial, particularly for difficult queries where traditional approaches may fail. On TREC Robust 2004, \texttt{QDER} doubles the relevant documents in top-10 positions compared to strong baselines and achieves a 78.5\% improvement in positioning highly relevant content.

Our work suggests that neural IR systems may need to rethink how query-document relationships are processed--moving away from static representations towards query-aware ones. \texttt{QDER}'s strong performance on challenging queries---those requiring understanding of multiple aspects or complex relationships---marks significant progress toward IR systems capable of handling real-world information needs.  The improvement in average rank for highly relevant documents (from 588.52 to 138.48) demonstrates our approach's effectiveness in identifying relevant content. Overall, we believe that future work can build on these insights by exploring deeper integration of query-aware mechanisms and more sophisticated entity-aware approaches to further enhance ranking effectiveness.

\bibliographystyle{ACM-Reference-Format}
\balance
\bibliography{references}
\end{document}